\documentclass[prl,twocolumn,showpacs,superscriptaddress]{revtex4-1}
\usepackage{graphicx,amssymb,amsmath,psfrag,color}
\usepackage[colorlinks=true,citecolor=blue,linkcolor=blue,urlcolor=blue]{hyperref}
\begin{document}
\definecolor{darkgreen}{rgb}{0,0.5,0}
\newcommand{\be}{\begin{equation}}
\newcommand{\ee}{\end{equation}}
\newcommand{\jav}[1]{\textcolor{red}{#1}}
%\bibliographystyle{apsrev4-1}
%\newcommand{\jav}[1]{#1}

%\title{Momentum-space mutual information in Luttinger liquids at finite temperature}
%\title{Disentangling thermal effects and entanglement in Luttinger liquids}
%\title{Distilling momentum space entanglement in Luttinger liquids at finite temperature}
%\title{Distilling momentum space entanglement in thermal Luttinger liquids}
\title{Distilling momentum-space entanglement in Luttinger liquids at finite temperature}

\author{Bal\'azs D\'ora}
%\author{authors}
\email{dora@eik.bme.hu}
\affiliation{Department of Theoretical Physics and MTA-BME Lend\"{u}let Spintronics Research Group (PROSPIN), Budapest University of Technology and Economics, 1521 Budapest, Hungary}
\author{Izabella Lovas}
\affiliation{Department of Theoretical Physics and BME-MTA Exotic  Quantum  Phases Research Group, Budapest University of Technology and
  Economics, 1521 Budapest, Hungary}
%\affiliation{Department of Theoretical Physics, Budapest University of Technology and Economics, 1521 Budapest, Hungary}
\author{Frank Pollmann}
\affiliation{Max-Planck-Institut f\"ur Physik komplexer Systeme, 01187 Dresden, Germany}

\date{\today}

\begin{abstract}
While much is known about the entanglement characteristics of ground states, the properties of reduced thermal density matrices have received significantly less attention.
Here we investigate the entanglement content  of reduced thermal density matrices for momentum-space bipartitioning in Luttinger liquids using analytical and numerical methods.
The low lying part of its spectrum contains an ``entanglement gap'', which persists up to temperatures comparable to the level spacing.
With  increasing temperature, the low energy modes acquire dispersion and resemble to those in the physical Hamiltonian with an enhanced effective temperature.
The momentum-space entanglement is carried by high energy modes (compared to temperature), featuring a completely flat spectrum. 
The von-Neumann entropy increases with temperature with a universal Sommerfeld coefficient. 
The momentum-space entanglement Hamiltonian turns out to be as universal as the physical Hamiltonian.

\end{abstract}


\maketitle

\paragraph{Introduction. }

Entanglement plays a major role in many distinct fields of physics. It is an essential tool
to characterize exotic phases of quantum matter, unveil topological and non-topological phases, describe the thermodynamics of black holes\cite{srednicki}, detect phase
transitions
without reference to any particular order parameter, classify various protocols in quantum
communication and computation\cite{nielsen}, design efficient numerical algorithms to attack strongly interacting systems etc.

Ground states of correlated systems described by local Hamiltonians are usually only slightly entangled.
This is inferred  from the area law scaling of their entanglement entropy from \emph{spatial} bipartitioning \cite{eisert,amico}.
%So where is the entanglement from the ground state?
By using momentum-space partitioning as opposed to real space partitioning\cite{thomale,qi,fuji,Mondragon,Andrade,Balasubramanian,PandoZayas2015,Diptarka},
ample entanglement scaling with the volume has been reported already for the ground states\cite{lundgren,doralundgren,berganza}.
Momentum-space partitioning in many cases is the most natural choice as interaction effects manifest themselves
more pronouncedly there. For example, non-interacting fermions carry spatial entanglement, while their momentum-space
wavefunction is a product state, i.e., a Slater determinant with no momentum-space entanglement. Non-zero entanglement only appears once interactions are switched on.

Another essential difference between real and momentum-space partitioning is that real space Hamiltonians
\`a la Hubbard model usually contain \emph{local} terms
 and are naturally \emph{homogeneous} due to translational invariance. Their momentum-space versions are \emph{non-local} and 
inherently \emph{inhomogeneous}
due to the separation of energy scales already at the non-interacting level.

Most entanglement studies have focused on $T=0$ ground states.
In reality, however, no quantum system is isolated perfectly from 
its environment\cite{Islam2015,junli,dai} and  experiments  are inevitably  conducted at finite temperature, therefore
understanding thermal effects on entanglement measures are vital.
In a mixed state, correlation characteristics are influenced by both thermal fluctuations \cite{arnesen} and entanglement (i.e. quantum non-local correlations)  effects and  distilling the latter  from the former is an important issue.
Addressing this is also relevant for quantum computation for controlling errors caused by to thermal excitations.

Thus motivated, we have studied  \emph{momentum-space} entanglement in 
interacting 1D systems with low energy excitation at thermal equilibrium, namely
thermal Luttinger liquids (LLs). 
While the spatial entanglement in 1D systems is well studied\cite{lefevre,drut}, 
much less attention has been given to the
entanglement in momentum-space partitionings.
We pose and answer several important questions: 
How do entanglement and thermal effects mix up in momentum-space partitionings of LLs?
What are the properties of the entanglement Hamiltonian\cite{lihaldane} at finite temperature? 
What happens to the entanglement gap \cite{thomale}?
How universal is the finite temperature entanglement spectrum?
%Which states are the source of entanglement?
Do we reach the thermodynamic entropy with increasing temperature?
Thanks to the 1D setting, analytical results can be obtained using bosonization and are backed up by numerics using exact diagonalization.
We find that the entanglement Hamiltonian of the right moving excitations is as universal as the full physical Hamiltonian, in spite of 
the non-trivial mixing of high and low energy modes.
It accounts successfully for the 
low energy excitation as  evidenced by focusing on the entanglement gap and the von-Neumann entropy.
This is in contract to fragility of topological entanglement at finite temperature\cite{castelnovo}.

\paragraph{Luttinger and interacting tight binding model} The low energy dynamics of LLs is  given in terms of bosonic sound-like collective excitations
with the Hamiltonian as\cite{giamarchi}
\begin{equation}
H=\sum_{q\neq 0}  v|q| b_q^{+} b^{\phantom{+}}_q
+\frac{g(q)}{2}[b^{\phantom{+}}_qb^{\phantom{+}}_{-q}+b_q^{+} b_{-q}^{+} ],
\label{ham0}
\end{equation}
where $b_q$ is the annihilation operator of a bosonic density wave, $g(q)=g_2|q|$, with $g_2$  the interaction strength, and $v$ the sound velocity of the non-interacting system.
Instead of $g_2$, the interaction is characterized by the dimensionless Luttinger parameter, $K$, %~\footnote{$K=1$ for the non-interacting case, and $K\gtrless 1$ for attraction/repulsion, respectively.},
 which is given for Eq. \eqref{ham0} by $K=\sqrt{(v-g_2)/(v+g_2)}$.
Eq. \eqref{ham0} is diagonalized by a Bogoliubov rotation, and the dispersion relation is $\omega_q=v_f|q|$
with the renormalized velocity $v_f=\sqrt{v^2-g_2^2}$.
The transformation gives the new boson operators as
$B_q=u_qb_q+v_qb^+_{-q}$ 
with Bogoliubov coefficients $u_q=\frac{K+1}{2\sqrt K}$ and $v_q=\frac{K-1}{2\sqrt K}$, which are $q$ independent for the present case.

Probably the best known lattice model, whose low energy dynamics is given by Eq. \eqref{ham0}, is a tight binding chain of one-dimensional
spinless fermions with nearest-neighbour interactions\cite{giamarchi,nersesyan}, which maps onto 
the 1D XXZ Heisenberg model by a Jordan-Wigner transformation\cite{giamarchi}.
In momentum-space, the Hamiltonian is
\begin{gather}
H=J\sum_{k}\cos(k)c_k^{\dagger} c_k^{\phantom{\dagger}}+\frac{J_z}{N}\sum_{k,p,q}\cos(q)c^\dagger_{p-q}c^{\phantom{\dagger}}_pc^\dagger_{k+q}c^{\phantom{\dagger}}_k,
\label{hmom}
\end{gather}
where $c$'s are fermionic annihilation operators in momentum-space, $N$ the number of lattice sites and $k=2\pi m/N$ for periodic boundary conditions, 
$m=1\dots N$. Its low energy physics is accounted 
for by Eq. \eqref{ham0} with %$v_f=\frac{\pi}{2}\frac{\sqrt{1-(J_z/J)^2}}{\arccos(J_z/J)}$ and 
$v_f={\pi}{\sqrt{1-(J_z/J)^2}}/2~{\arccos(J_z/J)}$ and
 $K=\pi/2[\pi-\arccos(J_z/J)]$, covering a wide range of $K$'s.

\paragraph{Reduced density matrix.}

At a given temperature $T$, the system is in a mixed state and its properties are described by the thermal
density matrix as
%\begin{gather}
$\rho(T)=\exp\left(-\beta H\right),$
%\end{gather}
where $\beta=1/T$. 
Instead of calculating the entanglement properties by partitioning our system in real space, we use the natural partitioning of Eq. \eqref{ham0} in terms
of right ($q>0$) and left ($q<0$) moving excitations and trace out all the left-movers. 
This is achieved by realizing that the bosonic operators for each $(q,-q)$ pair
 are the generators\cite{dorapdf} of the SU(1,1) Lie algebra~\footnote{The commutation relation is $[K_+(q),K_-(q)]=-2K_0(q)$, $[K_0(q),K_\pm(q)]=\pm K_\pm(q)$.}
 as $K_0(q)=(b^+_qb_{q}+b_{-q}b^+_{-q})/{2}$, $K_+(q)=b^+_qb^+_{-q}$, $K_-(q)=b_qb_{-q}$.
Then, the density matrix is rewritten following Ref. \cite{truax} as
\begin{gather}
\rho(T)=\prod_{q> 0}\exp[A_-(q) K_-(q)]\exp[2A_0(q) K_0(q)]\times\nonumber\\
\times\exp[A_+(q)K_+(q)],
\label{su11}
\end{gather}
and the $A_{\pm,0}(q)$ coefficients can be determined using Ref. \cite{truax}, but will not be needed for our purposes.

Since the contribution of distinct $(q,-q)$ pairs factorizes, it suffices  to focus on one of them.
By expanding the first and third exponentials in Taylor series,
the partial trace over the left-movers can be performed to yield $\sum_{n=0}^\infty a(n,q) (b_q)^n\exp[A_0(q) b^+_qb_q](b^+_q)^n$ for
a given momentum pair. This preserves the number of bosons with momentum $q$ as it commutes with the occupation number  $b^+_{q}b_{q}$.
Therefore, it is diagonal in occupation number basis and all of its (diagonal) matrix elements can be calculated, while the non-diagonal ones vanish.
As a result, we find that the reduced thermal density matrix for the right-movers is
\begin{gather}
\rho_{A}={\exp(-H_E)}/Z_E, \hspace*{5mm} H_E=\sum_{q>0}\varepsilon_q b^+_qb_{q},
\label{rhoA}
\end{gather}
where $H_E$ is the entanglement Hamiltonian, the single particle entanglement 
spectrum,  $\varepsilon_q$ can be determined from $A_{\pm,0}$ in Eq. \eqref{su11} and $Z_E=\prod_{q>0}(1-\exp(-\varepsilon_q))^{-1}$.

The entanglement spectrum is also determined by realizing that the reduced density matrix, $\rho_{A}$ accounts for expectation values of right-movers at all 
temperatures equally well  as the full $\rho(T)$. 
Since both are quadratic in the boson operators, it suffices to calculate
only the $\langle b^+_qb_{q}\rangle$ thermal expectation value from both density matrices, which defines immediately $\varepsilon_q$. 
Using this train of thoughts,
we get 
\begin{gather}
\varepsilon_q=\ln\left(\frac{|u_q|^2\exp(\beta\omega_q)+|v_q|^2}{|v_q|^2\exp(\beta\omega_q)+|u_q|^2}\right).
\end{gather}
The very same result is obtained by determining $\varepsilon_q$ from a direct calculation from Eq. \eqref{su11}.
It has two important limits: at $T=0$, it gives a completely flat, dispersionless spectrum and reduces to the universal result obtained in Ref. \cite{doralundgren}
as
$\varepsilon_q=\log((K+1)^2/(K-1)^2)$.
In the high temperature, $\omega_q\ll T$ limit, expanding the exponential yields a dispersive spectrum as 
$\varepsilon_q=\beta\omega_q/(|u_q|^2+|v_q|^2)=\beta\omega_q2K/(K^2+1)$. It resembles to the  spectrum of the physical Hamiltonian from Eq. \eqref{ham0}
at an effective temperature $T_{eff}=T(K^2+1)/2K$, which is always enhanced with respect to $T$.
The renormalization factor agrees with the anomalous exponent of Green's function of right movers\cite{giamarchi}.

\begin{figure}[h!]
\centering
\psfrag{x}[t][][1][0]{$J_z/J$}
\psfrag{z}[b][][1][0]{$\Delta_{EG}$}%{Entanglement gap}
\psfrag{y}[t][][1][0]{$J/TN$}
%\psfrag{yy}[b][t][1][0]{$S_{equilibrium}/S_{quench}$}
%\psfrag{t1}[t][][1.1][40]{equilibrium}
%\psfrag{t2}[t][][1.1][35]{sudden quench}
%\includegraphics[width=7cm]{EGscalingXXZ.eps}
\includegraphics[width=7cm]{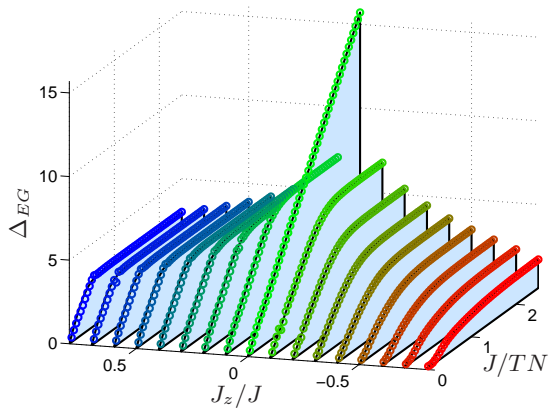}
\caption{(Color online) Entanglement gap at finite temperature for Eq. \eqref{hmom} using ED (circles) and for the Luttinger model  (light blue shaded areas) from 
Eq.  \eqref{EG} without any fitting parameter. The dimensionless temperature on the lattice, $TN/J$ is identified with $TL/v$ in bosonization.}
\label{EGfig}
\end{figure}

The $\varepsilon_q$'s are the single particle eigenvalues of the entanglement Hamiltonian at temperature $T$. Due to the bosonic
nature of excitations in Eq. \eqref{rhoA}, the many-body eigenvalues are built as $\sum_{q>0}n_q\varepsilon_q$, where the $n_q$'s are integers.
This would also allow to calculate the distribution function\cite{dorapdf} of 
momentum-space entanglement eigenvalues, similarly to Ref. \cite{lefevre}, what we leave for future work.

\paragraph{Entanglement gap.}

It is well established\cite{nersesyan,giamarchi} that the universal low energy dynamics of Eq. \eqref{hmom} is described by the Luttinger model in Eq. \eqref{ham0}.
We extend this universality to the entanglement Hamiltonian by focusing on its low energy excitations, in particular on the entanglement gap (EG).
The EG is  the difference between the ground state entanglement level and its first excited state. While at zero temperature, it carries unique
information about the entanglement properties of the system\cite{lihaldane}, at finite temperature entanglement and thermal effect get mixed and the entanglement
gap reflect their mutual influence on the eigenvalues of the entanglement Hamiltonian. Keeping this mixing in mind, we still keep on referring to it as EG.
The EG is thus corresponds to the the minimal single particle energy
as
\begin{gather}
\Delta_{EG}=\min_q \varepsilon_q=\varepsilon_{2\pi/L}=\nonumber\\
=\ln\left[\frac{(K^2+1)\coth(\beta v_f\pi/L)+2K}{(K^2+1)\coth(\beta v_f\pi/L)-2K}\right]
\label{EG}
\end{gather}
with $L$ the system size. This relation follows from the fact that 
in a finite size bosonized system, the excitation spectrum is sound-like and the minimal excitation energy occurs at momentum $2\pi/L$.
Above this gap, a continuum of many body entanglement levels occurs.
At $T=0$, it gives a finite EG, which vanishes with increasing temperature as
$\sim 2\pi v_f /LT_{eff}$.

The EG in LL from Eq. \eqref{EG} is compared to that obtained numerically for the interacting spinless tight binding chain using exact diagonalization (ED).
We consider half-filled systems with periodic boundary conditions and $N=6$, 10, 14, and  18 sites with non-degenerate, unique ground
states. Finite size scaling is performed to reach the thermodynamic limit (TDL).
\footnote{The XXZ Hamiltonian is projected to half filling, and its thermal density matrix is evaluated numerically at an initial temperature
$T=20J/N$. 
Projecting it back to the full system, the partial trace
over the left movers is performed, and the resulting matrix is diagonalized numerically. From its spectrum, we obtain the EG in 
the TDL after finite size scaling. Then, by multiplying the initial thermal density matrix by itself, we obtain the
density matrix at temperature $20J/N2$, and repeat the whole procedure. Subsequent multiplication with the initial thermal density matrix $n$ times
gives us access to the behaviour of the system at $20J/Nn$, without having to exponentiate the Hamiltonian.}
The comparison between LL and numerics is shown in Fig. \ref{EGfig} with convincing agreement.
For no interaction ($K=1$), the EG grows linearly with inverse temperature down to $T=0$, while in the presence of interaction, 
it saturates to its finite, $T=0$ value\cite{doralundgren} for $T<J/N$.

Now we comment on the universality of the momentum-space entanglement spectrum\cite{chandran,thomale,lihaldane}.
At $T=0$, the momentum-space entanglement Hamiltonian features a finite EG throughout the LL phase\cite{doralundgren,lundgren,lundgrenladder,berganza}, which is independent of
the system size.
This is in contrast to the spectrum of Eq. \eqref{ham0}, which becomes gapless in the TDL.
Increasing the temperature  enhances  the similarity between the two spectra as
the low energy modes, $\omega_q\ll T$ of the entanglement Hamiltonian become dispersive and gapless again, and behave similarly to the spectrum
of the physical Hamiltonian. The entanglement spectrum of high energy modes differs significantly from those in the original model and
remains flat, similarly to the  $T=0$ entanglement spectrum.

%Therefore, the low/high energy modes (compared to temperature) in the entanglement Hamiltonian are more thermal/entanglement like.

\paragraph{von-Neumann entropy and mutual information.}
In general, by dividing our system into a subsystem $A$ and the rest $B$, the amount of entanglement between $A$ and $B$ at zero temperature is excellently
characterized by the von Neumann or R\'enyi entropies of the reduced density matrix  of $A$.
However, at finite temperature, thermal effects contribute and influence the entanglement entropy\cite{korepin}.
Then, it is more useful the calculate the mutual information\cite{wolf,bernigau},
defined as
\begin{gather}
I(A,B)=S_A+S_B-S_{A\cup B},
\end{gather}
where $S=-\textmd{Tr}\rho\ln(\rho)$ is calculated from the respective finite temperature (reduced) density matrices. 
This measures entanglement at finite temperature by reducing thermal effects.

We start by recalling their behaviour in the case of \emph{spatial} bipartitioning for 1D critical systems.
The von-Neumann entropy satisfies an area law and depends on the logarithm of the size of the subregion at $T=0$, and crosses over to the conventional
thermodynamic entropy\cite{korepin} with increasing temperature with a volume law.
On the other hand, the log-scaling of mutual information  in the size of the subregion, inherited from the von-Neumann entropy, 
turns into a logarithmic dependence on the inverse temperature\cite{bernigau} as the temperature is enhanced.

In our case, $A$ ($B$)  consists of right- (left-) moving particles, while $A\cup B$ is the full system, and $S_A=S_B$.
Within the framework of bosonization,  both the reduced, $\rho_A$ and full, $\rho(T)$ density matrices are quadratic in terms of bosons, and their entropy follows from
\begin{gather}
S_C=\sum_{q\in C}\frac{\Omega_q}{\exp(\Omega_q)-1}-\ln\left(1-\exp(-\Omega_q)\right),
\label{entropy}
\end{gather}
where for the right moving partition ($C=A$), $\Omega_q=\varepsilon_q$, while for the whole system ($C=A\cup B$), $\Omega_q=\beta\omega_q$.
The momentum sum  in Eq. \eqref{entropy} indicates that a volume law holds for the momentum-space entropy for all temperatures.

For the full system, it is well-known\cite{giamarchi,nersesyan} that
the thermodynamic entropy in a LL behaves as $S(T)=\pi LT/3v_f$.
The coefficient of the term linear in $T$ is the Sommerfeld coefficient of the specific heat, $\gamma=\frac 1L\frac{\partial S}{\partial T}$, which even in the 
presence of interaction depends only on the velocity and is independent
from the LL parameter $K$, namely $\gamma_{LL}=\pi/3v_F$.
For the reduced density matrix of the right movers, however, we get
\begin{gather}
S_A=S_0+\frac{LT\gamma_{LL}}{2}+
\frac{LT}{2\pi v_f}\left[\ln\left(\frac{U}{V}\right)\ln\left(\frac{U+V}{UV}\right)\right.+\nonumber\\
+\left(1-3U\right)\textmd{Li}_2\left(-\frac 1V\right)-\textmd{Li}_2\left(\frac 1U\right)+\nonumber\\
+\left.
U\textmd{Li}_2\left(-\frac{U+V}{V^2}\right)
-\frac 12\textmd{Li}_2\left(\frac{V^2}{U^2}\right)\right]
\label{SaT}
\end{gather}
where $U=(K+1)^2/4K$, $V=(K-1)^2/4K$, Li$_2(x)$ is Spence's dilogarithm function.
It contains three different terms:  $S_0$ is the entanglement entropy at $T=0$, which is non-universal and satisfies a volume law. 
It is calculated from 
Eq. \eqref{entropy} with $\Omega_q=\ln((K+1)^2/(K-1)^2)$, and has already been analyzed\cite{doralundgren,lundgren,lundgrenladder,berganza}.
The second term is the entropy of a chiral LL, that is half of a conventional LL.
The third term is another contribution to the Sommerfeld coefficient, which, 
in sharp contrast to the thermal entropy, depends also on the LL parameter. This coefficient is 
universal and is determined solely by the low energy degrees of freedom of the theory.

\begin{figure}[h!]
\centering
\psfrag{x}[t][][1][0]{$J_z/J$}
\psfrag{y}[b][t][1][0]{\color{blue} $v\gamma_A$, \color{red} $v\gamma_{LL}$}
\psfrag{xx}[t][][1][0]{$J_z/J$}
\psfrag{yy}[t][b][1][0]{$v\partial I(A,B)/\partial T /L$}
\includegraphics[width=6cm]{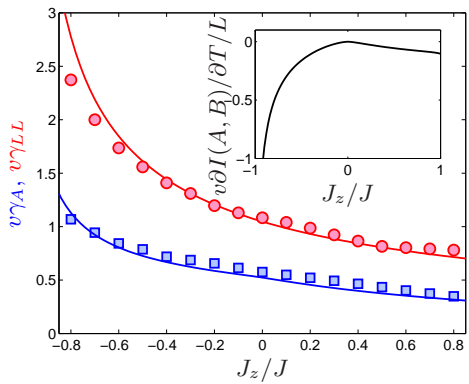}
\caption{(Color online) The Sommerfeld coefficients of the thermodynamic (red) and right-moving entanglement (blue) entropies for the interacting spinless fermion model (symbols)  and for the Luttinger model (solid curves) with no fitting parameter.
The inset visualizes the linear in $T$ coefficient of the mutual information.}
\label{sommerfeldfig}
\end{figure}

The Sommerfeld coefficient of the thermodynamic entropy of the interacting spinless or XXZ
Heisenberg model is known to follow $\gamma_{LL}$ at low temperatures, $T<J$, though the numerical data for $T<J/N$ is
dominated by finite size effects. Within this temperature window, we evaluate $S/T$ numerically and associate the
 Sommerfeld coefficient to its maximum.
Compared to   $\gamma_ {LL}$,  the agreement is convincing as shown in Fig. \ref{sommerfeldfig}.
 Next, we use the same method to extract the Sommerfeld coefficient
from $S_A$, and compare it with Eq. \eqref{SaT}. The agreement is again remarkable~\footnote{Had we identified the maximum in $\partial S/\partial T$ with
the Sommerfeld coefficient, qualitatively similar agreement would be obtained.}. 
The temperature dependent term appears for $T>J/N$, but $S_A$ does not approach thermodynamic entropy for $T<J$.

The mutual information starts from $2S_0$ at $T=0$ since the thermodynamic entropy vanishes identically in this limit. With increasing temperature,
it decreases linearly with the temperature with a universal coefficient, shown in  Fig. \ref{sommerfeldfig}. It only approaches zero for
$T\gg J$, which falls outside of the realm of the low energy description.
Extracting the linear in $T$ coefficient of the mutual information from ED is difficult due to finite size effects and reaching bigger systems would be beneficial\cite{melko}.
Nevertheless, the Luttinger model prediction is expected to be completely reliable, and demonstrates the importance of analytical methods in addressing entanglement in TDL.
For repulsive interactions, the mutual information acquires a very weak temperature dependence, as seen in the inset of 
Fig. \ref{sommerfeldfig}, and retains most of the entanglement of the $T=0$ ground state.
Attractive interaction reduce entanglement more efficiently and temperature effects are more pronounced.
%This could be due to the fact, that the $J_z\rightarrow -J$ state is very close to a classical phase separated state, while the
% $J_z\rightarrow J$ limit drives the system towards antiferromagnetism, which is far from being a classical N\'eel state.

The momentum-space mutual information satisfies a volume law as opposed to its real space version\cite{wolf}.
This is rooted back to the  difference between non-local and local Hamiltonians in momentum and real space, respectively.

\paragraph{Discussion.}

The entanglement between right- and left-moving excitations in a thermal Luttinger liquid is investigated.
The momentum-space entanglement Hamiltonian, determined analytically, is as universal as its full partner and accounts successfully for the low energy
excitations. The lowest levels feature an entanglement gap, which collapses with increasing temperature.
The momentum-space entanglement is carried by high energy modes (compared to temperature), featuring a completely flat entanglement spectrum.
The low energy modes resemble those in the full Hamiltonian and disperse linearly with momentum at enhanced effective temperature.
The von-Neumann entropy increases with temperature with a universal Sommerfeld coefficient and does not approach the thermodynamic entropy within the realm of the low energy theory, it only does so when
the temperature becomes at least comparable to the bandwidth.

Hamiltonians and correlations functions are naturally connected in real and momentum space by Fourier transformation.
Whether  such connection between entanglement entropies from real and momentum-space partitioning exists at all
is a future project and intriguing open question related to this research.
Besides being an interesting theoretical problem, numerical ramifications including the momentum-space DMRG algorithms\cite{ehlers,motruk} are also of high importance as well as
it can provides avenues for clever design of entangled states for quantum computation. 
Our current steps are crucial for revealing the structure in momentum-space entanglement.

\begin{acknowledgments}

We are grateful for the computing resources in the MPI-PKS and CPFS 
in Dresden. This research is supported by the Hungarian Scientific  Research Funds Nos. K105149, K108676, SNN118028
and by National Research, Development and Innovation Office – NKFIH K119442.
\end{acknowledgments}

\bibliographystyle{apsrev}
\bibliography{wboson1}

\end{document}